\documentclass{scrartcl}

\usepackage{graphicx}
\usepackage{url}

\def\VXLab{\textrm V\kern-.2em\lower.2em\hbox{\fontsize{.75em}{0em}\selectfont X}\kern-.1em{}L\kern-.35em\lower-.15em\hbox{\fontsize{.75em}{0em}\selectfont A}\kern-.15em B}

\begin{document}
\title{Towards a Cloud-based Architecture for Visualization and Augmented Reality to Support Collaboration in Manufacturing Automation}
\author{Ian D. Peake, Jan Olaf Blech, Shyam Nath, \\ Jacob Jacky Aharon, Argyll McGhie}
\date{RMIT University}
\maketitle
\begin{abstract}
In this report, we present our work in visualization and augmented reality technologies supporting collaboration in manufacturing automation.
Our approach is based on
(i) analysis based on spatial models of automation environments,
(ii) next-generation controllers based on single board computers,
(iii) cloud-, service- and web-based technologies
and
(iv) an emphasis on experimental development using real automation equipment.
The contribution of this paper is the documentation of two new demonstrators, one for distributed viewing of 3D scans of factory environments, and another for real time augmented reality display of the status of a manufacturing plant, each based on technologies under development in our lab and in particular applied to a mini-factory.
\end{abstract}

\section{Introduction}

Service-intensive cyber-physical systems can be expected to multiply in many domains and demand new service-based approaches to their maintenance.
Software- and service-intensive automation continues to rise but with perennial and increasingly urgent demands for improved system quality.
Traditionally, industrial automation occurs in standard-regulated and safety-critical automation environments such as nuclear reactors, ships or factories,
based on designs embodied in control circuits,
and more recently in many domains such as home, vehicle and building automation through programmable logic controllers.
Increasingly automation will apply more visibly to so-called cyber-physical systems in the consumer domain through robotic appliances
such as vacuum cleaners, telepresence devices, toys, drones.
Services to monitor, manage or service such devices will be required.
Multiple skill sets will be needed to bring new products and services economically through the full life cycle from cradle to grave.
These in turn can be expected to bring increased demand for technological aids to productivity.

The context of our work is to provide an architecture which
supports multiple online and possibly cloud-based services
including spatial analysis and visualization to support
engineering collaboration.
Spatial analysis provides the prospect of system-level
insight into questions of consistency through simulation at
more meaningful levels of abstraction based on design models.
Support of this kind is in line with
goals of trends such as Industry 4.0,
increasing system intelligence by the use of high
level domain-specific models.
In our work we focus on support for live decision making and collaboration at critical moments of service or maintenance phases~\cite{BPS+15},
in particular decision support and collaboration focuses on spatial analysis and insight about automated processes
such as would be expected to occur in a running factory environment.

In this paper we focus on visualization of real time factory
data by 3D scanning and feedback via augmented reality.
By 3D scanning we are referring to depth or LIDAR-style scanning which provides a
3D point cloud representation of a space.
Such scanning is provided in our experimentation by low cost Microsoft XBox Kinect(TM) sensors
which combine an optical and depth camera.
In a factory setting, and given the possibility of mechanical failures,
complex failures may occur that
cannot be detected using existing sensors,
or cannot be properly rectified without deep insight into the
internal software-configured behaviour of the system.
3D scanning, while requiring new computation resources,
has the potential to
provide richer and more computationally-accessible
information about the state of
factory components than would be provided by typical sensors
and actuators, or even standard video cameras.

Augmented reality (AR) aims to provide additional sensory information to a user which
intuitively enriches their experience or appreciation of their environment.
Augmentation may involve any of the senses; in this project we focus on visual augmentation.
In a factory environment, AR has the potential to improve the productivity of staff during commissioning and down time
by providing better operator guidance.
In our work we aim in particular to improve two aspects of software engineering:
first, to provide more practical and meaningful interpretations of software component output,
all of which might otherwise require complex and specialised interpretation of trace output;
second, to simplify collaborative debugging
by making data jointly visible to local and remote technical staff.

\begin{figure}
\centering
\includegraphics[width=1.0\textwidth]{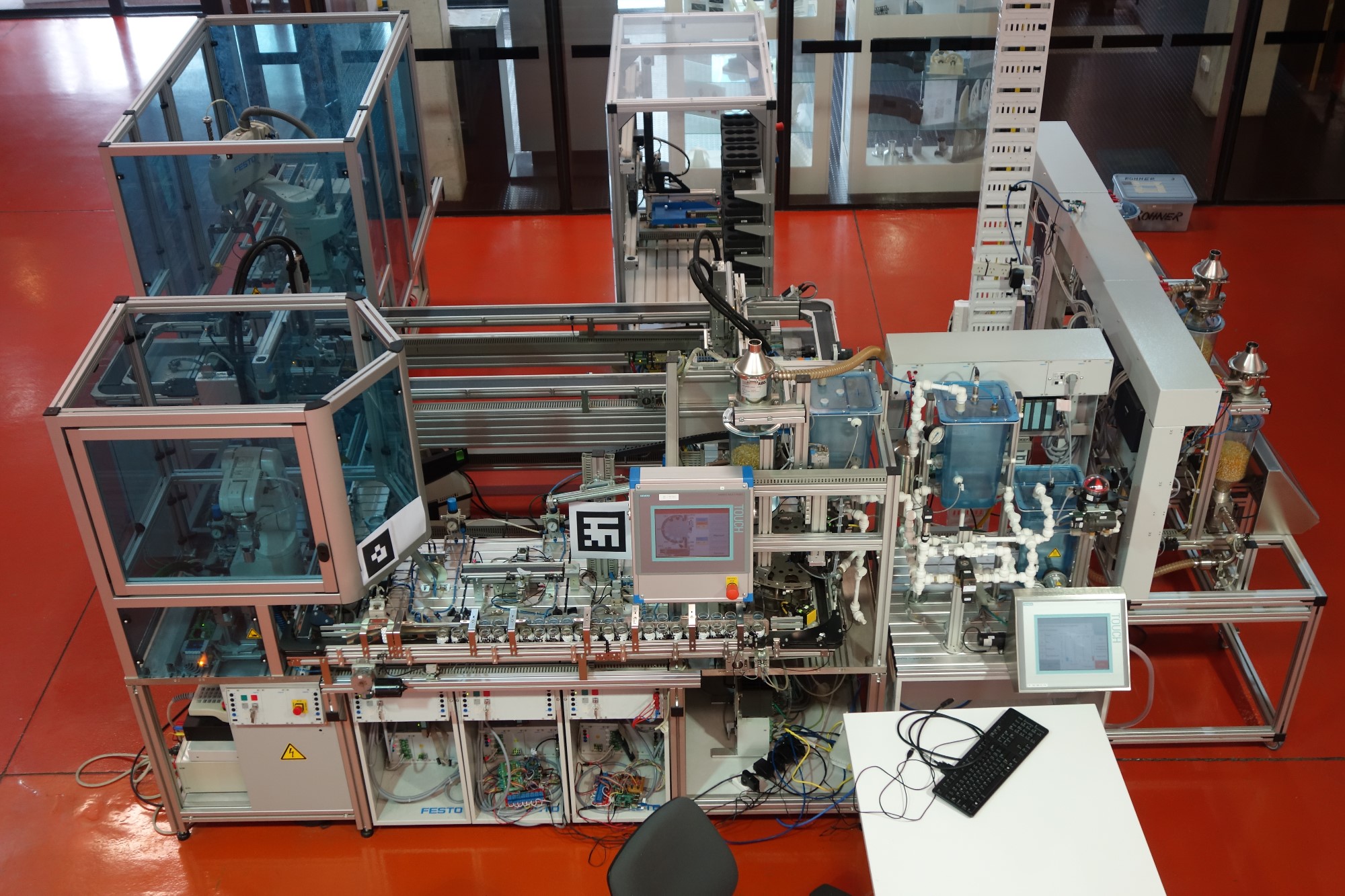}
\caption{Food processing factory demonstrator}
\label{fig:factory}
\end{figure}

Our Virtual Experiences Laboratory (\VXLab{} ~\cite{vxlab-2015}) was designed as a global distributed lab to explore collaborative engineering with specific attention to visualization and monitoring supported by spatial analysis.
Within \VXLab{}, our Global Operations Visualization laboratory (GOV Lab) has been developed as a prototyping platform for exploring visualization and collaboration for management and control of remote distributed software-intensive industrial automation facilities.
A food processing factory demonstrator (shown in Figure~\ref{fig:factory}) from FESTO provides an experimental platform for case studies,
with custom controllers developed around commodity single board computers.
Among other work, we developed a concept demonstration for an augmented reality view of the factory demonstrator although not based on live data~\cite{etfa-workshop-AR-paper-2016}. A cloud-based monitoring infrastructure aimed at the factory demonstrator is described in~\cite{PB17}.

\begin{figure*}
\includegraphics[width=1.0\textwidth]{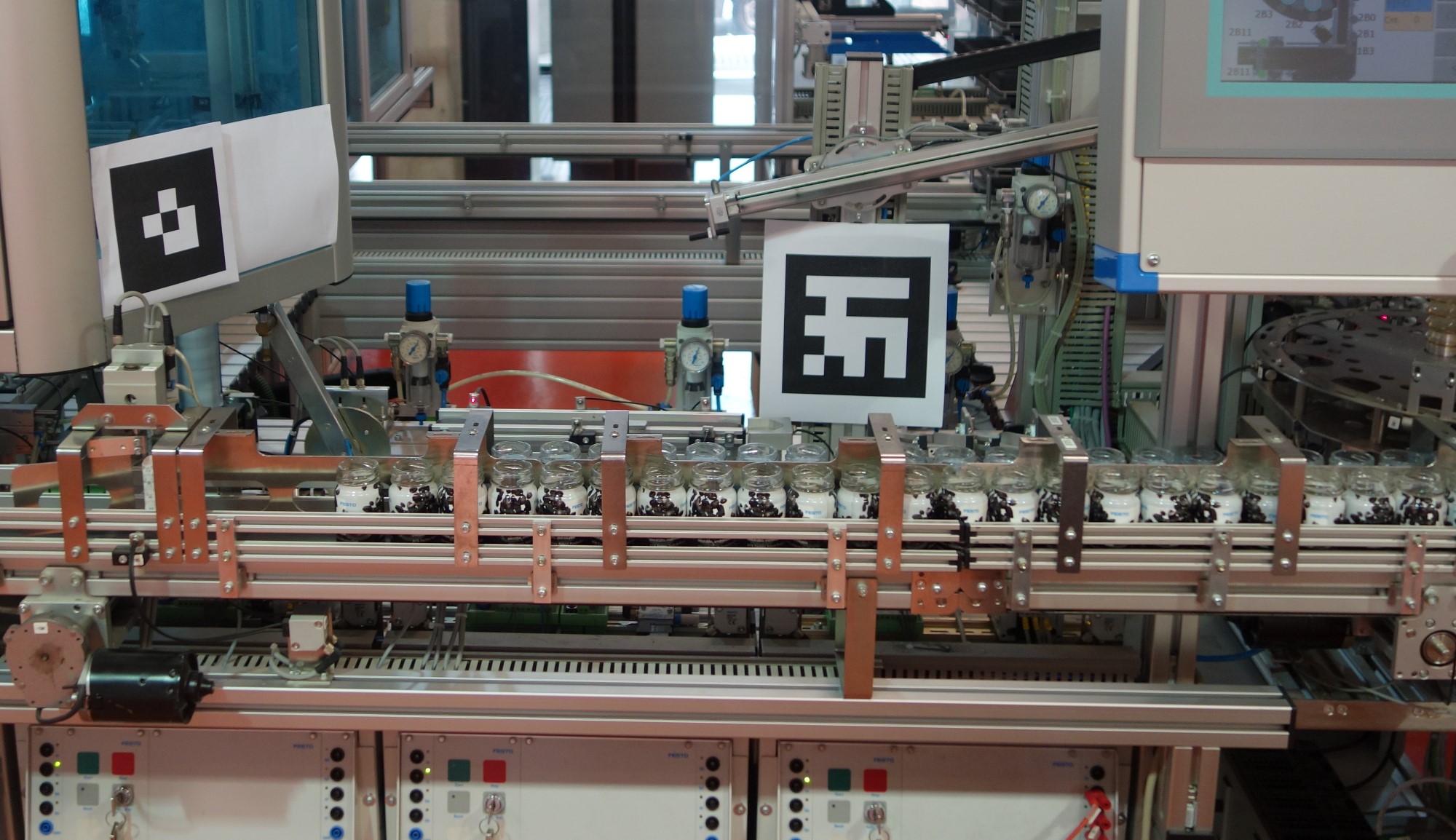}
\caption{Factory demonstrator - cap handling stage}
\label{fig:factory-caps}
\end{figure*}

An example use case for our technology is in spatial measurement.
For example in our packaging plant, bottles are filled with solid (corn kernels) and/or liquid (water) products.
Bottles are carried in groups of six in pallets, which are transferred on conveyors.
Bottles are picked from pallets, a robot uncaps bottles,
then caps are transferred for recapping.
In the cap transfer stage (shown in Figure ~\ref{fig:factory-caps}),
a piston pushes caps from the bottom of a stack,
a pneumatic suction-based pick-and-place unit places caps on a conveyor,
from the conveyor caps are placed one by one onto a staging area
to be picked up by a sliding picking and capping unit for placing back onto bottles.
In this paper for simplicity we focus on a failure where
the pick-and-place unit may fail to grip a cap,
and consider how a sensor which scans a wide area may be able to
detect that the cap has not been picked up, or indeed dropped
and fallen off the line.
Another major potential advantage of 3D sensing in this context is
to detect such a failure among a large number of possible failures in
a number of nearby factory components.

In this work we describe several extensions of \VXLab{} which have been developed, as follows:
GOV Lab has been extended to include a spatial visualization service capability,
displaying three dimensional (point cloud) views of automation processes such as factories.
The factory demonstrator has been extended to provide Scala-based code for the custom controllers,
which interfaces with sensors and actuators, and provides real time data from selected sensors to a cloud-based real time spatial
monitoring service based on the BeSpaceD framework.
Finally an augmented reality system based on mobile devices has been developed which displays live notifications overlaid on
the factory demonstrator based on an interpretation of sensor data.

The contribution of this paper is the documentation of the combination of experimental extensions above,
and its relation to underlying research and supporting architecture for spatial analysis and cloud-based collaborative engineering support.

\section{Architecture}

\begin{figure*}
\includegraphics[width=1.0\textwidth]{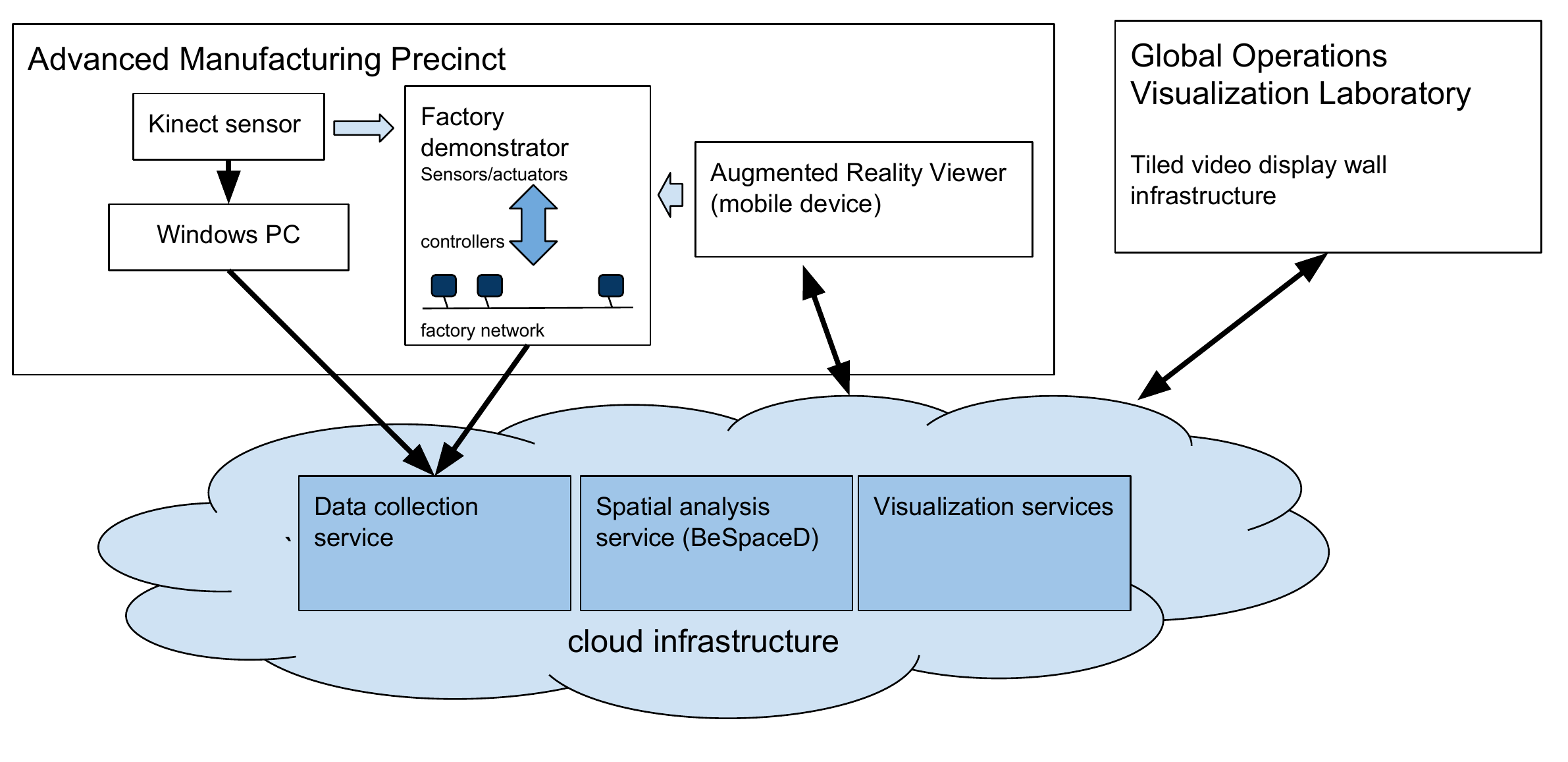}
\caption{Architecture}
\label{fig:arch}
\end{figure*}

\begin{figure}
\centering
\includegraphics[width=1.0\textwidth]{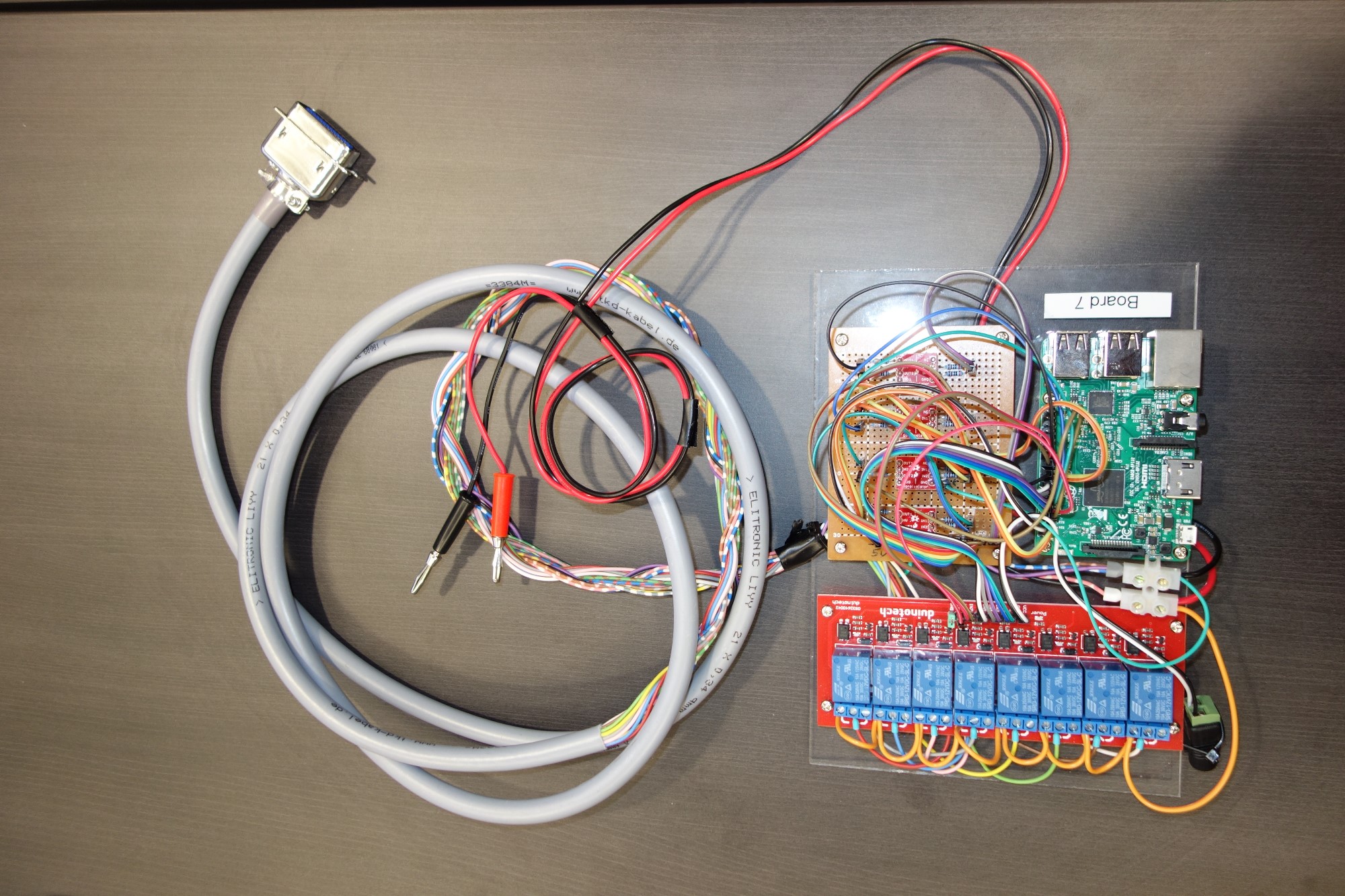}
\caption{Experimental controller based on single board computer}
\label{fig:pi-controller}
\end{figure}

Our architecture is as shown in Figure~\ref{fig:arch}.
The factory demonstrator in RMIT's advanced manufacturing precinct (AMP) is equipped with a Kinect sensor and viewed with a mobile device with a rear-facing camera and web browser.
The Kinect sensor is supported by a Windows PC which scans the environment and transforms the sensor data into point cloud form and uploads data to cloud-based visualization services.
Custom controllers based on Raspberry Pi single board computers
(shown in Figure~\ref{fig:pi-controller})
are connected to sensors and actuators related to cap transfer,
and upload sensor data via message queueing protocols to a cloud-based monitoring service.
Sensor data is processed by a BeSpaceD-based spatial analysis service
and a NodeJS-based visualization service.
Sensor or point cloud data is then retrieved from visualization services.
The GOV Lab (separate location approx 500m from AMP) provides a monitoring
room with a large tiled visualization wall used for displaying video feeds
and custom visualization.

The GOV Lab uses the Scalable Amplified Graphics Environment SAGE2~\cite{sage2},
a web-based service framework providing
large virtual shared desktops
for use by multiple users and multiple geographically distributed sites.
Although SAGE2 ``scales down'' to a single web browser it is best suited to large
tiled display walls, where multiple web browsers on separate display segments are
combined virtually to provide a single logical display surface suitable for display
of custom applications.
In particular the Euclidian point cloud visualization application developed at RMIT is
based on the SAGE2 framework and can therefore be viewed using the GOV Lab.
For the purposes of demonstration and evaluation,
services are hosted in the
the Australian National eResearch Collaboration Tools and Resources (Nectar) cloud,
which is a national distributed cloud computing infrastructure for Australian public research
and teaching institutions.
To improve bandwidth for video monitoring, reduce cost and latency,
the GOV Lab's SAGE2 server is typically hosted in an internal private network,
however, during evaluation, for displaying the point cloud data using Euclidian,
the GOV Lab was successfully connected to a SAGE2 server running in the
Nectar cloud.

Architectural concerns include standard compliance, message reliability and end-to-end latency.

\section{Spatial Modeling and Analysis}

BeSpaceD~\cite{bespaced,bespaced0} is integrated into the system providing potential support for spatial modeling and analysis.
The raw sensor data is collected from the Raspberry Pi-based controller board,
which has a lightweight version of the BeSpaceD language.
The Raspberry Pi transforms the data into BeSpaceD objects,
using a BeSpaceD Scala model
which impregnates the spatial data onto the raw sensors data,
and communicates that object
into the Nectar hosted Ubuntu server for analysis, using AMQP queue protocol, specifically using
RabbitMQ JAVA libraries. Once the server received the incoming message it will employ the
BeSpaceD engine to analyse the current state of Festo, considering not only the raw sensors data,
but the spatial data within, considering all the relevant previous messages, and then output current
state via web-sockets, to the augmented reality client, so that the current state of Festo could be shown on screen.
In our demonstration, we only used one source of information for the analysis, which is the RPI and
the cap missing event.

\section{Distributed Immersive Visualization}

We aim at providing distributed immersive visualizations, prototyping a tool for interactively viewing and measuring point clouds in the SAGE2 environment.

Our previous work has assumed the existence of high quality video streams broadcast / shared between multiple control centres using
the Scalable Adaptive Graphics Environment (SAGE) and its successor SAGE2.
Here we take a step towards providing three dimensional, explorable representations of the process of interest in the form of point clouds.

\begin{figure*}
\includegraphics[width=1.0\textwidth]{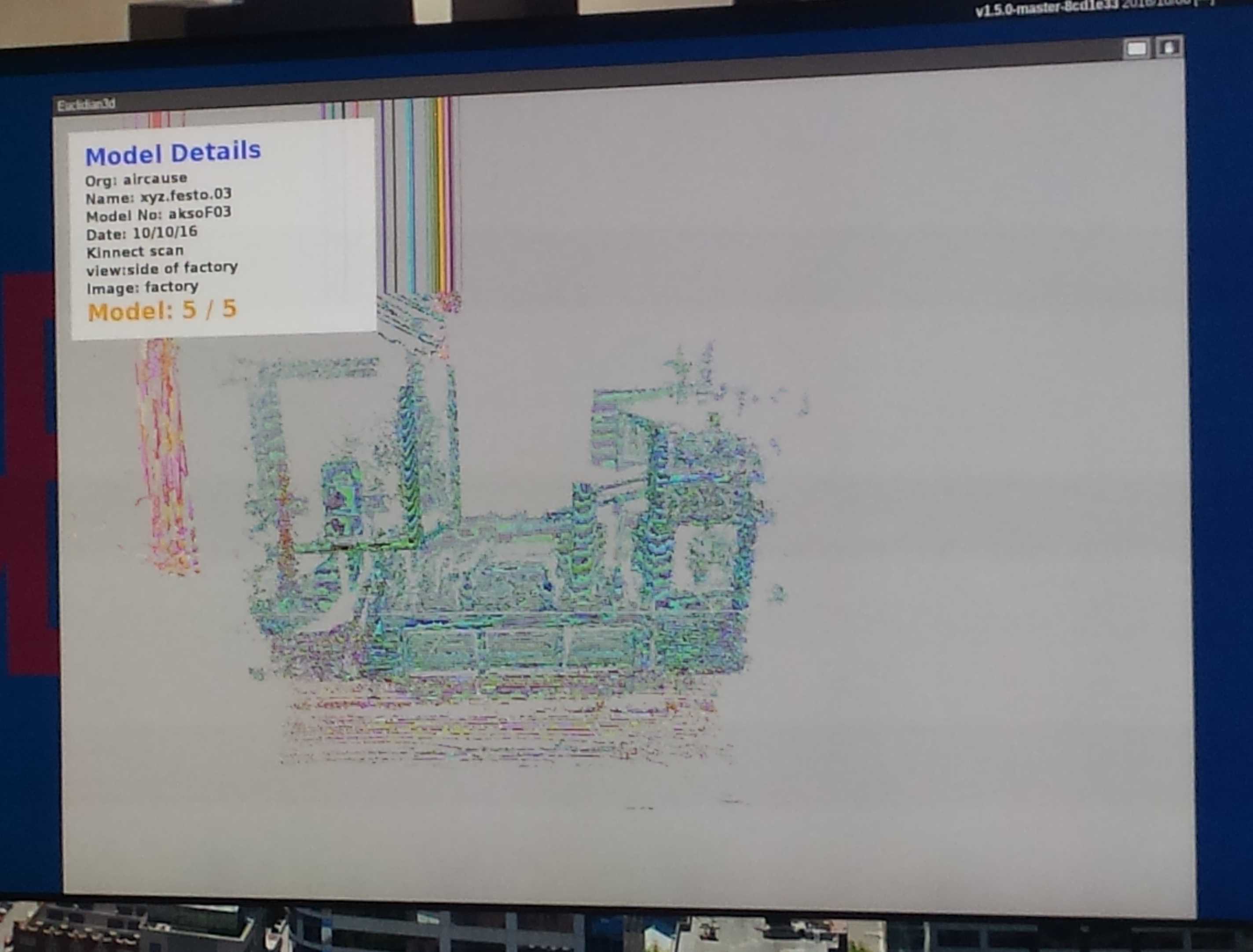}
\caption{Point cloud viewer showing 3D scan of factory demonstrator}
\label{fig:pointcloud-viewer}
\end{figure*}

Figure~\ref{fig:pointcloud-viewer} shows the output of Euclidian\footnote{For a video of the Euclidian point cloud viewer application, see~\url{https://www.youtube.com/watch?v=om3L9AzNC-M}},
a point cloud viewer developed for use in GOV Lab on the SAGE2 platform\footnote{sage2.sagecommons.org}.
Point clouds are generated via a Microsoft Xbox Kinect time-of-flight depth sensor.
The Kinect generates a rectangular array of scalar values representing the nearest points in an approximately rectangular grid of rays projected into a scene, or alternatively, a sample of the surface as seen from the sensor's point of view.
A point cloud represents the 3D position of each spatial point.
Currently a single Kinect scan has been generated,
thus the point cloud may be considered a representation of a surface visible from one point of view.
Using the 3D point cloud representation has the advantage that the additional depth information can be used to improve analysis over what could be done using video alone: such analysis could be both automatic, or manually by humans.
For humans the benefit of 3D point cloud data can be realised for example in the form of a capability
to change the point of view, thus to explore
the scene in an intuitive way.
Where colour information is not available, Euclidian uses an algorithm to assign colours to enhance the presentation of depth from a three-dimensional perspective.

\begin{figure*}
\includegraphics[width=1.0\textwidth]{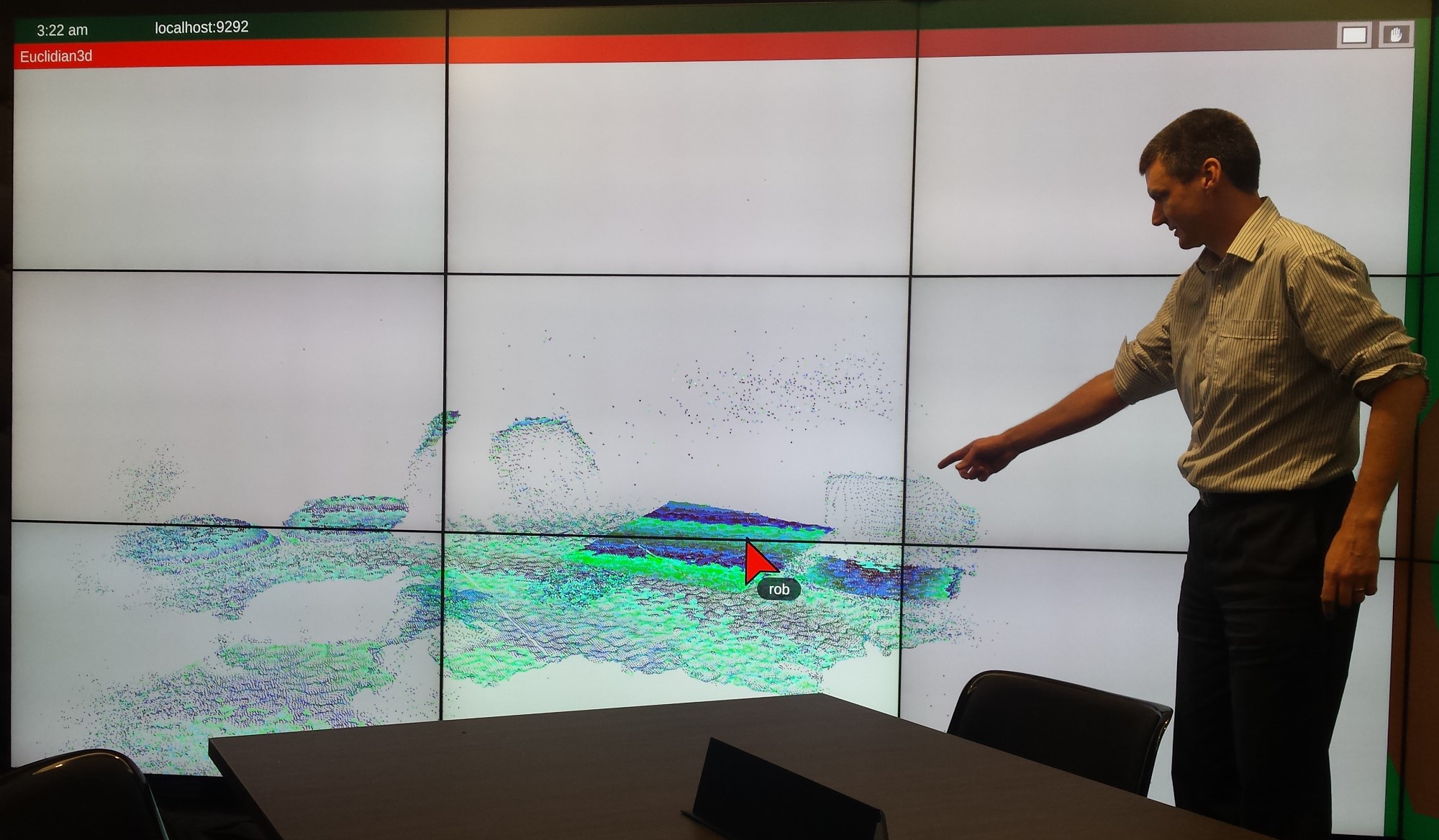}
\caption{Point cloud viewer in GOV Lab}
\label{fig:pointcloud-viewer-govlab}
\end{figure*}

Euclidian provides typical common actions used in a 3D scene such as camera movement (pan/zoom).
Euclidian's keyboard and mouse-driven control interface allows the user to intuitively interact with three-dimensional point-cloud models by standard camera operations such as panning and zooming.
In addition it provides a measuring box feature which enables users to size and align the box manually with respect
to clusters of interest within a scene.
As a SAGE2 application,
Euclidian provides two additional features, namely multi-location real time display and displayability on high resolution tiled display walls.
As a web-based technology, the SAGE2 framework provides mechanisms to enable multiple users to view and interact with applications.
For instance, a public SAGE2 service can be accessed by multiple display clients.
Alternatively, SAGE2 services can be linked, and applications running on a given SAGE2 service can be shared with another, such that their
visible state are linked, and interactions via either service are synchronised.
For example, Figure~\ref{fig:pointcloud-viewer-govlab} shows a user viewing a point cloud on our tiled video wall in GOV Lab,
while a remote user in another city controls the interface.

\begin{figure*}
\includegraphics[width=0.95\textwidth]{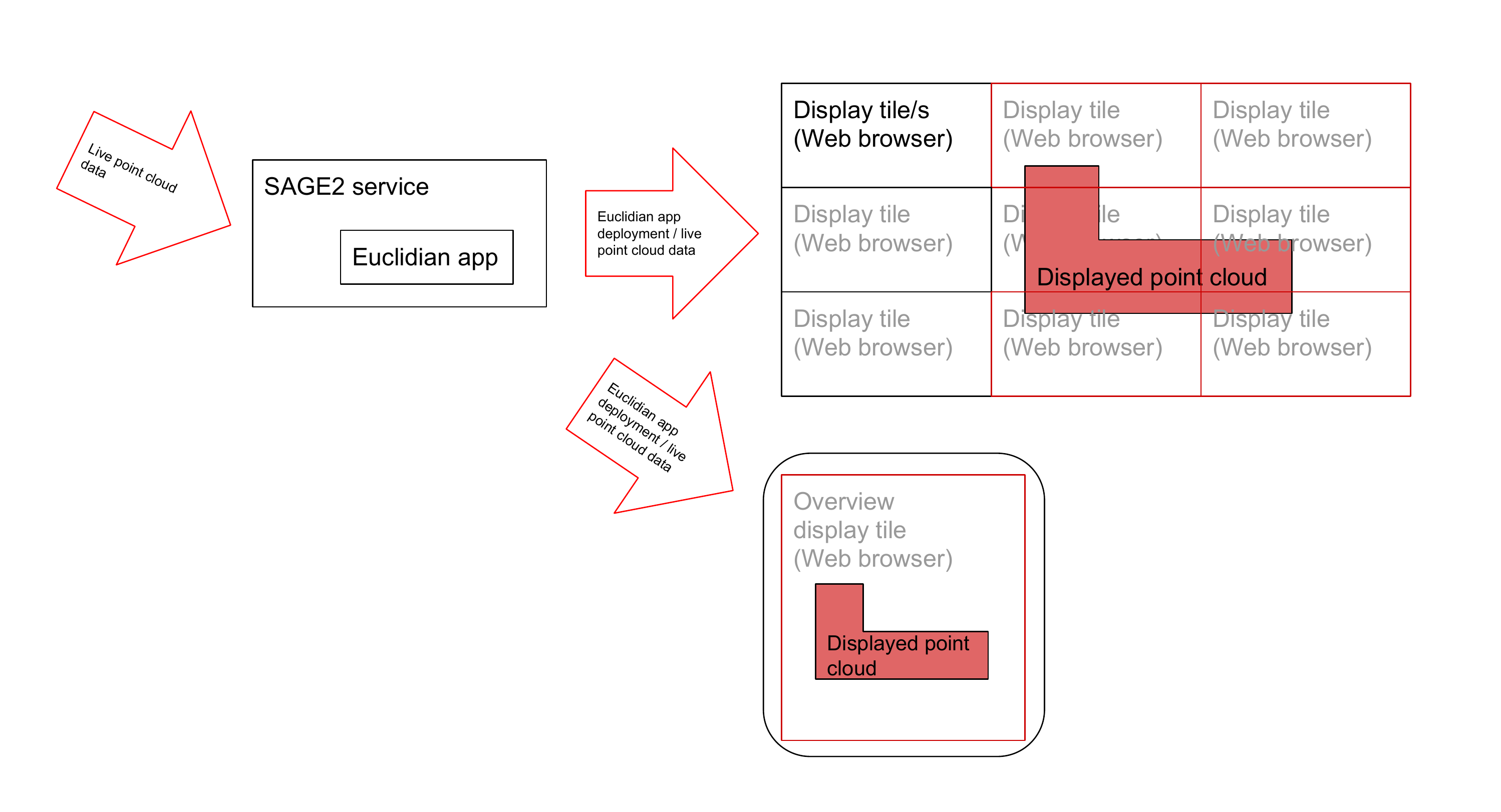}
\caption{Euclidian architecture (with SAGE2)}
\label{fig:euclidian-architecture}
\end{figure*}

Figure~\ref{fig:euclidian-architecture} shows the architecture of the
Euclidian SAGE2 application,
based on a standard SAGE2 model
for application distribution and event updates to displays.
The SAGE2 service deploys the Euclidian application written in Java\-script
to all relevant displays, that is, to all web browsers which are
displaying some portion of a Euclidian application.
There may be multiple browsers redundantly displaying content from the same
logical display tile, for example in multiple geographical locations.
Individual Euclidian display application instances are responsible
for clipping their own viewport into the 3D scene so as to
create a single coherent view on a tiled display.
This is achieved by extending a standard SAGE2 Java\-script class.
Depth sensor data in the form of update messages containing point clouds (set of x,y,z values) is received by the SAGE2 service using a special message type.
Sensor data is currently collected using a partial port of a websocket extension  used by SAGE2, called WebsocketIO.
The SAGE2 service relays data to the web browsers displaying some portion of a Euclidian; currently this is achieved using the
standard SAGE2 ``broadcast'' message\footnote{Live data updates is an experimental extension of the original Euclidian application}.
The deployment of Euclidian was facilitated by the use of Nectar cloud infrastructure, which provided reliable hosting of the Ubuntu server. The server made use of firewall-based accessibility for improved quality of service; by fortifying the server against security threats involving the denial-of-service, and also preventing general high-traffic network congestion. An independent web service enabled users to administer various aspects of the server, including the ability to configure the SAGE2 service.

Benchmarking of Euclidian was coordinated across multiple sites across Australia by multiple users;
synchronized testing was rigorously performed, and compared the individual clients.
Visual observation instrumentation of performance (frames per second)
suggests that with one or two users there are no noticeable delays.
With the introduction of additional users, there is a noticeable impact on updates reflected in frames per second measurable on remote clients.

\begin{figure*}
\includegraphics[width=1.0\textwidth]{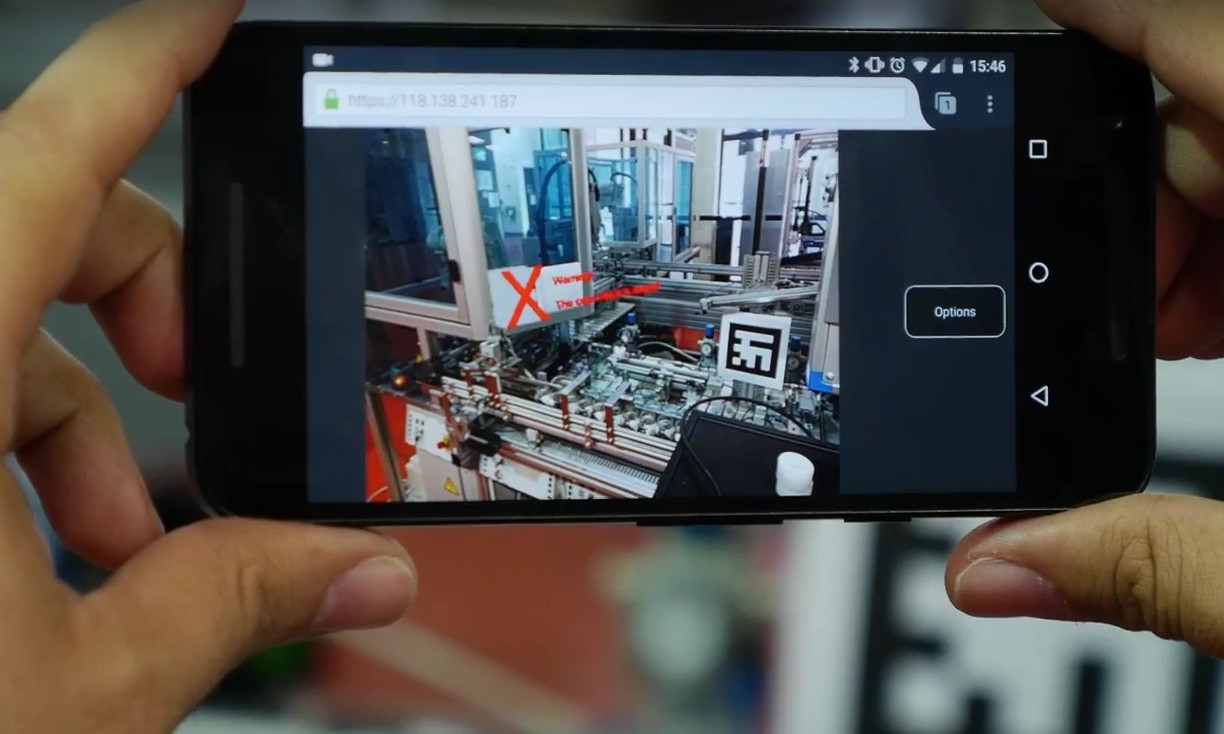}
\caption{Augmented reality view of factory demonstrator}
\label{fig:ar-mobile}
\end{figure*}

Point cloud data messages are currently encoded as simple JSON objects, for example:
\begin{verbatim}
{"values": [ "0.0 1.1 1.1","2.0 3.1 2.0",
	     "1.0 1.0 2.0", "1.0 3.0 2.0",
	     "1.0 0.0 1.0","0.5 1.0 1.0" ]}
\end{verbatim}
Such a representation has several potential disadvantages given the expected
bandwidth requirements of raw point cloud data and the real time, low latency
requirements.
In particular, use of JSON likely has a significant impact on the performance of the
Euclidian display JavaScript code, since there is a cost to parse (unstructured) JSON messages
which could likely be avoided with other representations.

\section{Augmented Reality}

Based on real-time data provided to cloud services,
we provide real-time and contextualised display of
information, via mobile devices, that might not otherwise be obvious to
an operator, even an experienced operator with deep knowledge of factory
control systems.
A demonstrator system has been developed which can overlay live information
over a view of the system on a mobile device.
Information is correlated with the physical location of the device
which would enable an operator to select physical factory elements
of interest and view information about these elements.
Currently the demonstrator indicates whether a stack of caps is empty
and perhaps requires manual intervention, or not, overlaid on the mobile
device view of the stack.
Figure~\ref{fig:ar-mobile} shows the demonstrator in use\footnote{A video of the augmented reality subsystgem is available at: \url{https://www.youtube.com/watch?v=ROtKYLeGF9I}}.

For the web augmented reality display we used JSARToolkit, an open sourced,
Java\-Script-based framework that
allows transposing a 3D object onto a 2D video as an overlay.
As for earlier \VXLab{} demonstrators, the viewer relies on fixed fiducial markers
to align a live video stream with a virtual 3D spatial view containing extra
information.
JSARToolkit is a port of ARToolkit which supports detection of markers and alignment~\cite{artoolkit}.
The web client holds all the 3D models within.
Once the Web/mobile client loads, it registers for incoming messages from the server.
The server pushes new control information, after BeSpaceD analysis, to the client.
Thus, only control messages need to be transferred via the network, making real time augmented reality achievable.

\section{Related work and evaluation}

Our work aims at the combination of virtual and augmented reality and industrial automation.
It connects with areas Industry 4.0~\cite{industrie40},
since we aim to use existing data sources and models from the demonstrator.
Applications of augmented reality to factory environments have been explored since the early 1990s~\cite{CM92}, along with the exploration of human factors~\cite{NUA98}.
Several authors have described similar recent work in the application of augmented reality to manufacturing~\cite{vrar-manufacturing-book13}.
Related work for industrial automation has also been studied~\cite{CON} in particular for location discovery for a similar factory demonstrator~\cite{SMBV}.

We have also experimented with using the underlying visualization platform, SAGE2, to provide immersive,
CAVE-like virtual reality for a Smart Grid visualization concept demonstrator SmartSpace 3D,
similar in style to what is achieved by Microsoft Research in their holoportal demonstrators~\cite{msr-holoportation},
using our {\em Virtual Experience Portals} (VxPortals)~\cite{vxportals-2016}\footnote{For video of the VxPortals and SmartSpace 3D application, see \url{https://www.youtube.com/watch?v=9MjCfZyyymI}}.

\section{Conclusion}

We have presented a software architecture integrating services for spatial scanning, visualization, augmented reality and analysis.
The systems described have demonstrated in a running factory environment.
All service elements have been evaluated in a cloud-based setting.
Similar and related services can be expected to become increasingly available at scale due to the ease of deployability.
However there are further challenges to commercial scalability,
in particular the scope for variation in various manufacturing domains
will make it challenging to integrate data such as factory models.

The underlying visualization platform SAGE2 is targeted primarily at flat, high resolution, tiled display walls.
We plan to extend the Euclidian point cloud viewer to view live streams of spatial data and to port Euclidian to the VxPortals based on our previous work.
A web-based rigid body simulation of the FESTO demonstrator has also been developed to augment the live data generated by the
actual physical factory demonstrator with the intention of exploring the use of simulation in conjunction with monitoring and visualization.

\end{document}